\documentclass[RNAAS]{aastex63}

\usepackage[]{natbib}
\setcitestyle{aysep={}}

\begin{document}

\title{Deconfusing the Confusogram: \\ Getting New Insights from Zeeman Doppler Imaging}

\correspondingauthor{J. Sebastian Pineda}
\email{sebastian.pineda@lasp.colorado.edu}

%\author{J. Sebastian Pined}
%\affiliation{}

%% The \author command can take an optional ORCID.

\author[0000-0002-4489-0135]{J.~Sebastian Pineda}
%\altaffiliation{\href{mailto:sebastian.pineda@lasp.colorado.edu}{sebastian.pineda@lasp.colorado.edu}}
\affiliation{University of Colorado Boulder, Laboratoy for Atmospheric and Space Physics, 3665 Discovery Drive, Boulder CO, 80303, USA}

\author{James. R. A. Davenport}
\affiliation{Astronomy Department, University of Washington, Box 951580, Seattle, WA 98195, USA}

\begin{abstract}
	Zeeman Doppler Imaging is a powerful tool for characterizing the strength and topology of stellar magnetic fields. In this research note, we present a new way to visualize the typical results from ZDI for an ensemble of stars, addressing some of the concerns with the standard `confusogram' approach to illustrating the data. Our publically available plotting methods further enable an accessible means to consider variability in the inferred magnetic field topologies from repeated observations, as we demonstrate with the literature ZDI data on M dwarfs.
\end{abstract}
%% Note that RNAAS manuscripts DO NOT have abstracts.

%% Start the main body of the article. If no sections in the 
%% research note leave the \section call blank to make the title.
\section{} 

Magnetic fields are responsible for a host of phenomena (e.g., starpspots, flares) that transform stars from relatively tranquil luminous orbs to eruptive and dynamic objects. %From an internal dynamo, the fields power non-thermal heating of the upper atmosphere, driving chromospheric and coronal emissions across the entire high-energy spectrum. The hot coronal plasma also drives stellar winds that control the angular momentum evolution. In addition to defining the stellar surface features from spots to plage, the emerging magnetic fields can erupt with bright flares and powerful mass ejections.
These aspects of stellar astrophysics have significant consequences for exoplanetary systems, especially around low-mass M-dwarfs which exhibit much stronger magnetic phenomena across their lifetimes than more massive stars. 

One of the most powerful tools to characterize the magnetic fields responsible for these processes is Zeeman Doppler Imaging \citep[ZDI;][]{Semel1989}. Although there are caveats given the inherent assumptions \cite[e.g., see within][]{Morin2008}, through time resolved observations of the polarized emission features in high-resolution spectra, ZDI enables a reconstruction of the likely surface topology of the stellar magnetic field. ZDI has thus successfully revealed the variety and complexity of magnetic fields in a multitude of objects \citep[see within][]{Vidotto2014b}.

The ZDI inferred stellar magnetic field (mainly the large-scale field) is represented by several quantities: the surface average field strength, total reconstructed magnetic energy, and the fraction of that energy in dipole, quadrapole and other modes. To understand trends in these topology metrics, these data are often displayed with the mass, and rotation period (or another age proxy) of each star. Relating all of the quantities of interest thus typically requires graphically displaying at least four or five variables in a single plot.

Faithfully representing these data thus presents a challenge within the standard representations of scientific articles. The solution in the literature has been plots that have been referred to as `confusograms' \citep{Vidotto2016}, see Figure~15 of \cite{Morin2010} for a traditional example or Figure~1 of \cite{See2016} for a more recent iteration. %Although the term may have additional meanings, within stellar astrophysics it generally refers to the ZDI plots. 
Rather than aid in their presentation, the term itself quickly elides the fact that the plots may actually obscure their underlying data, especially for astronomers not accustomed to making or viewing them. %, i.e., generally most anyone outside the ZDI community. 

If the visual data representation chosen is not clear, then there is something fundamentally amiss. The standard ZDI `confusogram' attempts to plot five variables at once by morphing the shape of individual data points from a five-pointed star to a decagon to convey the degree of axisymmetry, in addition to the x and y axes (rotation period and mass), the color of individual points (fraction of energy in poloidal fields), and their size (total magnetic energy). This symbol choice provides a clear mental connection from point shape to magnetic field symmetry, however, the perception of the shape change becomes mixed with the size of the points. This conflates the two parameters of field axisymmetry and total magnetic energy. %For example, two points with the same magnetic energy but one entirely axisymmetric and the other asymmetric should be the same size but the star shaped point would seem smaller as it takes up much less of the plotting area. 

\begin{figure}[hb]
	\centering
	\includegraphics[width=0.8\textwidth]{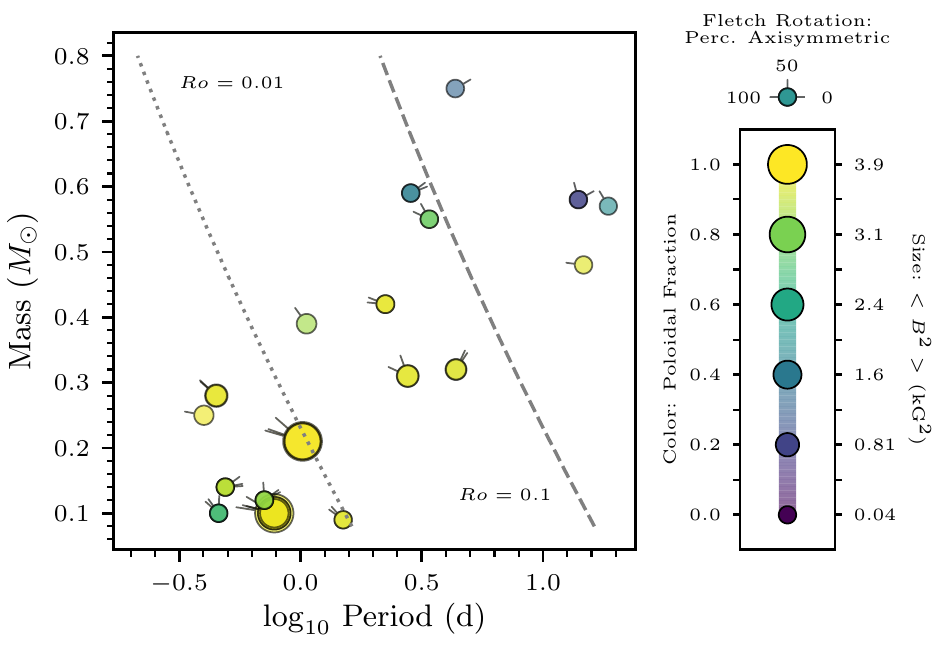} % requires the graphicx package
	\caption{The typical results inferred from Zeeman Doppler Imaging, displayed in a manner addressing many of the concerns of the `confusogram', and enabling comparisons of multiple epochs in a single plot. We present the mass and period of stars in the ZDI sample, shown with fletched scatter points whose rotation indicate degree of magnetic field axisymmetry (see legend), color corresponding to fraction of magnetic energy in poloidal field, and area indicative of total magnetic energy. Contours of constant Rossby number ($Ro = P / \tau_{c}$) from \cite{Wright2018} are indicated as dotted ($Ro = 0.01$), and dashed ($Ro = 0.1$) lines. Scatter points with multiple fletches show different epochs of the same star.}
	\label{fig}	
\end{figure}

We have endeavored to address some of these issues in the plotting of ZDI data, with our results shown in Figure~\ref{fig}, presenting the collected ZDI data of M dwarfs from literature sources \citep{Donati2008,Morin2008,Morin2010,Vidotto2014b}. We use circular points with variable areas for all data, and instead add a fletch (or tail) to each to indicate the degree of axisymmetry through the rotation of the symbol (see Figure~\ref{fig}). This allows the clearer separation of the field axisymmetry from total magnetic energy. 

This innovation enables us to represent an additional aspect of the data: the topologies from multiple epochs of the same star, usually averaged in the standard `confusogram'. Epoch changes in the axisymmetry can be seen through fletches in different directions. Using the plot point transparency, differences in the magnetic energy, are evident as overplotted concentric circles, and distinct poloidal fractions between epochs are revealed through the blended colors of the fill symbols. %The main literature result that more axisymmetric fields (left pointing fletches in Figure~\ref{fig}), correspond to largely poloidal topologies is well evident. 
Our view of the different ZDI epochs provides a means to visualize variability in the topology. This view of the data further prompts the question of whether fields with low poloidal fractions also show greater variations in the degree of axisymmetry, as some stars appear to illustrate (see Figure~\ref{fig}).

The visual representation of data can have a strong influence on their interpretation and possible scientific conclusions. We hope our representation is a step towards a more accessible means to digest the rich information available through ZDI, and have provided \texttt{python} code explicitly for this purpose.\footnote{DOI: \href{https://doi.org/10.5281/zenodo.3872224}{10.5281/zenodo.3872224}  } Examining time variability of magnetic field topologies will also enable new studies of the relation between these fields and other observable magnetic processes, and we encourage more long-term observations of these stars to reveal whether and to what extent their topologies do vary. While our code is designed with ZDI data in mind, it may be generally applicable for additional problems requiring the joint illustration of 5 or 6 variables in a single two-dimensional plot.

\acknowledgments

JRAD acknowledges support from the DIRAC Institute in the Department of Astronomy at the University of Washington. The DIRAC Institute is supported through generous gifts from the Charles and Lisa Simonyi Fund for Arts and Sciences, and the Washington Research Foundation. The authors would like to thank Marin Anderson and Melodie Kao for useful feedback in the development of our plotting methods.

%\begin{minipage}{\textwidth}

%\end{minipage}

%\bibliographystyle{aasjournal}
%\bibliography{../../../Documents/biblio.bib}

\begin{thebibliography}{}
	\expandafter\ifx\csname natexlab\endcsname\relax\def\natexlab#1{#1}\fi
	\providecommand{\url}[1]{\href{#1}{#1}}
	\providecommand{\dodoi}[1]{doi:~\href{http://doi.org/#1}{\nolinkurl{#1}}}
	\providecommand{\doeprint}[1]{\href{http://ascl.net/#1}{\nolinkurl{http://ascl.net/#1}}}
	\providecommand{\doarXiv}[1]{\href{https://arxiv.org/abs/#1}{\nolinkurl{https://arxiv.org/abs/#1}}}
	
	\bibitem[{{Donati} {et~al.}(2008){Donati}, {Morin}, {Petit}, {Delfosse},
		{Forveille}, {Auri{\`e}re}, {Cabanac}, {Dintrans}, {Fares}, {Gastine},
		{Jardine}, {Ligni{\`e}res}, {Paletou}, {Ramirez Velez}, \&
		{Th{\'e}ado}}]{Donati2008}
	{Donati}, J.-F., {Morin}, J., {Petit}, P., {et~al.} 2008, \mnras, 390, 545,
	\dodoi{10.1111/j.1365-2966.2008.13799.x}
	
	\bibitem[{{Morin} {et~al.}(2010){Morin}, {Donati}, {Petit}, {Delfosse},
		{Forveille}, \& {Jardine}}]{Morin2010}
	{Morin}, J., {Donati}, J.-F., {Petit}, P., {et~al.} 2010, \mnras, 407, 2269,
	\dodoi{10.1111/j.1365-2966.2010.17101.x}
	
	\bibitem[{{Morin} {et~al.}(2008){Morin}, {Donati}, {Petit}, {Delfosse},
		{Forveille}, {Albert}, {Auri{\`e}re}, {Cabanac}, {Dintrans}, {Fares},
		{Gastine}, {Jardine}, {Ligni{\`e}res}, {Paletou}, {Ramirez Velez}, \&
		{Th{\'e}ado}}]{Morin2008}
	{Morin}, J., {Donati}, J.~F., {Petit}, P., {et~al.} 2008, \mnras, 390, 567,
	\dodoi{10.1111/j.1365-2966.2008.13809.x}
	
	\bibitem[{{See} {et~al.}(2016){See}, {Jardine}, {Vidotto}, {Donati}, {Boro
			Saikia}, {Bouvier}, {Fares}, {Folsom}, {Gregory}, {Hussain}, {Jeffers},
		{Marsden}, {Morin}, {Moutou}, {do Nascimento}, {Petit}, \& {Waite}}]{See2016}
	{See}, V., {Jardine}, M., {Vidotto}, A.~A., {et~al.} 2016, \mnras, 462, 4442,
	\dodoi{10.1093/mnras/stw2010}
	
	\bibitem[{{Semel}(1989)}]{Semel1989}
	{Semel}, M. 1989, \aap, 225, 456
	
	\bibitem[{{Vidotto}(2016)}]{Vidotto2016}
	{Vidotto}, A.~A. 2016, in 19th Cambridge Workshop on Cool Stars, Stellar
	Systems, and the Sun (CS19), Cambridge Workshop on Cool Stars, Stellar
	Systems, and the Sun, 147
	
	\bibitem[{{Vidotto} {et~al.}(2014){Vidotto}, {Gregory}, {Jardine}, {Donati},
		{Petit}, {Morin}, {Folsom}, {Bouvier}, {Cameron}, {Hussain}, {Marsden},
		{Waite}, {Fares}, {Jeffers}, \& {do Nascimento}}]{Vidotto2014b}
	{Vidotto}, A.~A., {Gregory}, S.~G., {Jardine}, M., {et~al.} 2014, \mnras, 441,
	2361, \dodoi{10.1093/mnras/stu728}
	
	\bibitem[{{Wright} {et~al.}(2018){Wright}, {Newton}, {Williams}, {Drake}, \&
		{Yadav}}]{Wright2018}
	{Wright}, N.~J., {Newton}, E.~R., {Williams}, P. K.~G., {Drake}, J.~J., \&
	{Yadav}, R.~K. 2018, \mnras, 479, 2351, \dodoi{10.1093/mnras/sty1670}
	
\end{thebibliography}

\end{document}